\documentclass[letter]{aa} 

\usepackage{graphicx}
\usepackage{txfonts}
\usepackage{natbib}
\usepackage{soul}
\usepackage{longtable}
\usepackage{lscape}
\usepackage{threeparttable}
\usepackage{longtable}
\usepackage[titletoc]{appendix}
\bibpunct{(}{)}{;}{a}{}{,} 
\usepackage{color}

\begin{document} 
	\title{The heart of Sakurai's Object revealed by ALMA}
	
	
	\titlerunning{The heart of Sakurai's Object revealed by ALMA}
	\authorrunning{D.~Tafoya et al.}
	
	\author{Daniel~Tafoya\inst{1}, Peter~A.~M.~van Hoof\inst{2}, Jes\'{u}s~A.~Toal\'{a}\inst{3}, Griet Van de Steene\inst{2}, Suzanna
		Randall\inst{6}, Ramlal Unnikrishnan\inst{1}, Stefan Kimeswenger\inst{4,5}, Marcin Hajduk\inst{7}, Daniela Barr\'{i}a\inst{8} \and Albert
		Zijlstra\inst{9}}
	\institute{Department of Space, Earth and Environment, Chalmers University of Technology, Onsala Space Observatory, 439~92 Onsala, Sweden\\
		\email{daniel.tafoya@chalmers.se}
		\and
		Royal Observatory of Belgium, Ringlaan 3, 1180 Brussels, Belgium
		\and
		Instituto de Radioastronom\'{i}a y Astrof\'{i}sica, UNAM, Ant. carretera a P\'{a}tzcuaro 8701, Ex-Hda. San Jos\'{e} de la Huerta, Morelia 58089, Mich., Mexico
		\and
		Universit{\"a}t Innsbruck, Institut f{\"u}r Astro- und Teilchenphysik, Technikerstr. 25, 6020 Innsbruck, Austria 
		\and
		Universidad Cat\'olica del Norte, Instituto de Astronom{\'i}a, Av. Angamos 0610, Antofagasta, Chile
		\and European Southern Observatory, Karl Schwarzschild str. 2, 85748 Garching, Germany
		\and
		Space Radio-Diagnostics Research Centre, University of Warmia and Mazury, Prawocheńskiego 9, 10-720 Olsztyn, Poland
		\and
		Facultad de Ingeniería y Arquitectura, Universidad Central de Chile, Av. Francisco de Aguirre 0405, La Serena, Coquimbo, Chile
		\and
		Jodrell Bank Centre for Astrophysics, Alan Turing Building, University of Manchester, Manchester, M13 9PL, UK
	}
	\date{\today}
	
	
	\abstract{We present high angular-resolution observations of Sakurai's object using the Atacama Large Millimeter Array, shedding new light on its morpho-kinematical structure. The millimetre continuum emission, observed at an angular resolution of 20~milliarcsec (corresponding to 70 AU), reveals a bright compact central component whose spectral index indicates that it composed of amorphous carbon dust. Based on these findings, we conclude that this emission traces the previously suggested dust disc observed in mid-infrared observations. Therefore, our observations provide the first direct imaging of such a disc. The H$^{12}$CN($J$=4$\rightarrow$3) line emission, observed at an angular resolution of 300~milliarcsec (corresponding to 1000 AU), displays bipolar structure with a north-south velocity gradient. From the position-velocity diagram of this emission we identify the presence of an expanding disc and a bipolar molecular outflow. The inclination of the disc is determined to be $i$=72$^\circ$. The derived values for the de-projected expansion velocity and the radius of the disc are $v_{\rm exp}$=53~km s$^{-1}$ and $R$=277 AU, respectively. On the other hand, the de-projected expansion velocity of the bipolar outflow detected in the H$^{12}$CN($J$=4$\rightarrow$3) emission of approximately 1000~km~s$^{-1}$. We propose that the molecular outflow has an hourglass morphology with an opening angle of around 60$^{\circ}$. Our observations unambiguously show that an equatorial disc and bipolar outflows formed in Sakurai's object in less than 30 years after the born-again event occurred, providing important constraints for future modelling efforts of this phenomenon.}
	
	
	\maketitle
	
	
	\section{Introduction}
	
	Stars with masses up to 8~$M_\odot$ shed most of their hydrogen envelope during the Asymptotic Giant Branch (AGB) phase. As hydrogen and helium burning ceases in the layers around the core, they transition to the early white dwarf sequence where they spend the remainder of their lives as white dwarfs (WDs). Theory predicts that up to one-quarter of these stars experience a `final helium shell flash' during this stage, where the residual helium spontaneously ignites under degenerate conditions \citep{Iben1983,Bloecker2001,Herwig2005}, which triggers a so-called `born-again event'. During the born-again event, the remaining hydrogen envelope mixes into the stellar core and is burned, leading to a new H-poor and $^{13}$C-rich shell ejection. After this, the star undergoes a double loop in the Hertzsprung-Russell diagram \citep{Lawlor2003,Hajduk2005} before helium burning stops, and the star re-enters the WD sequence as an H-poor star. The born-again event is thought to be a likely path for the formation of hydrogen-deficient stars such as non-DA white dwarfs, PG 1159, and [WC] stars, as well as some R~CrB stars \citep{Werner2006}. Because born-again objects evolve on very short timescales, it is extremely rare to observe the process in action. In fact, this has been achieved for only two cases: V605 Aql, which experienced a final helium shell flash at the beginning of the twentieth century, and V4334 Sgr (also known as Sakurai's object), which was caught undergoing the born-again behaviour in the mid 1990's \citep{Duerbeck1996}. Consequently, the born-again event is still very poorly understood, despite its importance to stellar evolution and evolved stellar populations.
	
	Sakurai's object, whose assumed distance from the Sun is 3.5 kpc \citep{Hinkle2020}, is the only star to have been monitored with modern facilities throughout most of the born-again process. At the time of discovery, Sakurai's object had evolved from the early WD sequence to a born-again red giant just a few years earlier. Soon after that, it became completely enshrouded by a thick layer of dust \citep{Kimeswenger1997}. Early optical observations already provided indications that the morphology of the ejecta was bipolar with a thick obscured disc or torus \citep{Tyne2000, Kerber2002, Evans2006, vanHoof2007}. From MiD-Infrared/VLT Interferometer (MIDI/VLTI) observations it was possible to derive a model of the morphology of the ejecta and determine that it contains a dusty disc seen almost edge-on, efficiently screening the central source \citep{Chesneau2009}. In addition, Gemini Near InfraRed Imager/ALTtitude conjugate Adaptive optics for the InfraRed unit (NIRI/ALTAIR) and Near-Infrared Integral Field Spectrometer (NIFS) images of Sakurai's object show a pair of lobes expanding with a velocity of $\sim$300~km~s$^{-1}$, with the north-east and south-west lobes moving away from and toward us, respectively \citep{Hinkle2014,Hinkle2020}. While all these observations point towards some form of bipolar structure, they only give a partial and limited view of the situation. Since it is expected that a considerable fraction of the ejected material has cooled down and condensed into molecular gas and dust, observations at longer wavelengths are essential to getting the full picture. In this letter we present high angular-resolution Atacama Large Millimeter Array (ALMA) observations to uncover the morphology of the cool ejecta in the youngest born-again star.

	\section{Observations}\label{sec:obs}
	
	We conducted observations of the continuum and molecular line emissions of Sakurai's object using ALMA. The project IDs associated with the observations used in this work are 2017.1.00017.S, 2018.1.00088.S (PI: P. van Hoof), and 2018.1.00341.S (PI: D. Tafoya). All observations were performed using the ALMA 12~m array, which included 43-47 antennas. The relevant parameters of the observations are summarised in Table \ref{Tab:1}. The data were calibrated and imaged using the ALMA pipeline, which is included in the Common Astronomy Software Application \cite[CASA;][]{McMullin2007}.
	
\begin{figure}
		\begin{center} 
			\includegraphics[angle=0, width=\linewidth]{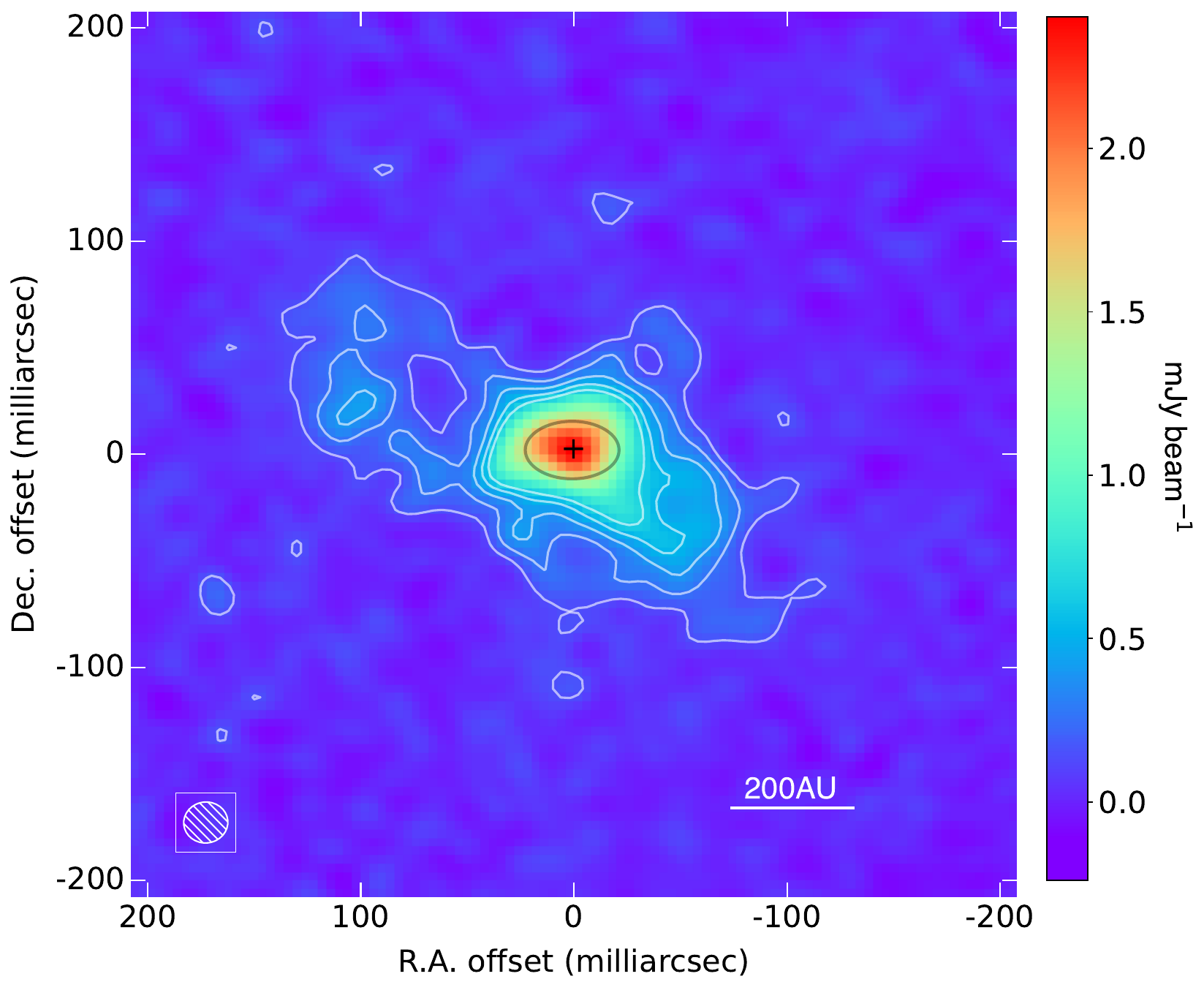}
			\caption{ALMA continuum emission of Sakurai's object at 233~GHz. The emitting regions consists of a bright compact central component and faint extended structures elongated in the northeast-southwest direction. A Gaussian fit to the central component is shown as a grey ellipse. The black cross indicates the continuum peak position at (J2000) R.A.=17$^{\rm h}$52$^{\rm m}$32$\rlap{.}^{\rm s}$6990 $\pm$ 0$\rlap{.}^{\rm s}$0002, Dec.=$-$17$^{\circ}$41$^{\prime}$7$\rlap{.}^{\prime\prime}$915 $\pm$ 0$\rlap{.}^{\prime\prime}$003. The rms noise level of the image is 44 $\mu$Jy beam$^{-1}$. The horizontal bar indicating the linear scale of the image assumes a distance of 3.5 kpc to the source. The synthesised beam of the ALMA observations is shown in the bottom-left corner and its parameters are: $\theta_{\rm beam}$$=$21$\times$19 milliarcsec, P.A.$=$$-$84.1$^{\circ}$.}\label{fig:Fig1}
		\end{center} 
	\end{figure}

	For the continuum emission, we included data from all the observations. However, for the molecular line emission, we only used the data from the observations of project 2018.1.00341.S. The continuum emission was obtained separately for each spectral window using line-free channels. The high angular-resolution continuum image, shown in Fig.~\ref{fig:Fig1}, was made only from the observations carried out on 22 June, 2019. The continuum emission from four spectral windows centred at 224.0~GHz, 226.0~GHz, 240.0~GHz, and 242.0~GHz was combined, resulting in a total bandwidth of 6.73~GHz. The multi-scale algorithm was used in the cleaning process with the CASA task `tclean'. The robust parameter of the Briggs weighting scheme and the pixel size of the image were set to 0.5 and 0.004 arcsecs, respectively. The resulting root mean square (rms) of the continuum image is approximately 44~$\mu$Jy~beam$^{-1}$, with a beam size of 0.021 $\times$ 0.019 arcsec (P.A.\footnote{The position angle (P.A.), measured north through east, relative to the north celestial pole.} $-$84.1$^{\circ}$).
	
	The spectral data includes emission of the H$^{12}$CN($J$=4$\rightarrow$3), H$^{13}$CN($J$=4$\rightarrow$3), and CO($J$=3$\rightarrow$2) lines. The spectral setup of these observations consists of four 1.875 GHz wide spectral windows centred at frequencies 343.009 GHz, 344.967 GHz, 355.009 GHz, and 356.897 GHz. Each spectral window has 1917 channels with a width of 976.562 kHz. The channel maps were created manually using a robust parameter of 0.0 and a pixel size of 0.06 arcsecs. The typical rms noise in the individual 0.8 km~s$^{-1}$ wide channels is 3 mJy beam$^{-1}$, with a beam size of approximately 0.33 $\times$ 0.30 arcsec (P.A. $\sim$73$^{\circ}$).

	\section{Results and discussion}
	\label{sec:results_discussion}
	
	\subsection{Continuum emission: The dust disc of Sakurai's object}
	
	The circumstellar dust of Sakurai's object has been extensively studied at IR wavelengths \cite[e.g.][and references therein]{Geballe2002,Tyne2002,Kaufl2003,Chesneau2009,Hinkle2014,Hinkle2020,Evans2020,Evans2022}. In contrast, observations of the dust at sub-millimetre or longer wavelengths remain limited, with only a few instances reported \cite[e.g.][]{Evans2004}. From an analysis of the evolution of the circumstellar dust in Sakurai's object over an $\sim$20 year period, \citet{Evans2020} find that, overall, the mid-IR dust emission can be adequately modelled as originating from a blackbody that has cooled from $T_{\rm dust}$=1200~K in 1998 to $T_{\rm dust}$=180~K in 2016. Additionally, the dust mass has increased from $M_{\rm dust}\sim10^{-9}~M_{\odot}$ to $M_{\rm dust}\sim10^{-5}~M_{\odot}$ in the same period. While the mid-IR emission indicates a decline of the dust temperature, the increase of the continuum emission in the near-IR emission suggests the formation of new hot dust \citep{Hinkle2020,Evans2020,Evans2022}. Thus, a multi-temperature model to fit both cooling, expanding dust and a hotter near-IR excess is required, and observations at millimetre (mm) and sub-millimetre (sub-mm) wavelengths are invaluable to constrain physical parameters.
	
	\begin{figure}
		\begin{center} 
			\includegraphics[angle=0,width=\linewidth]{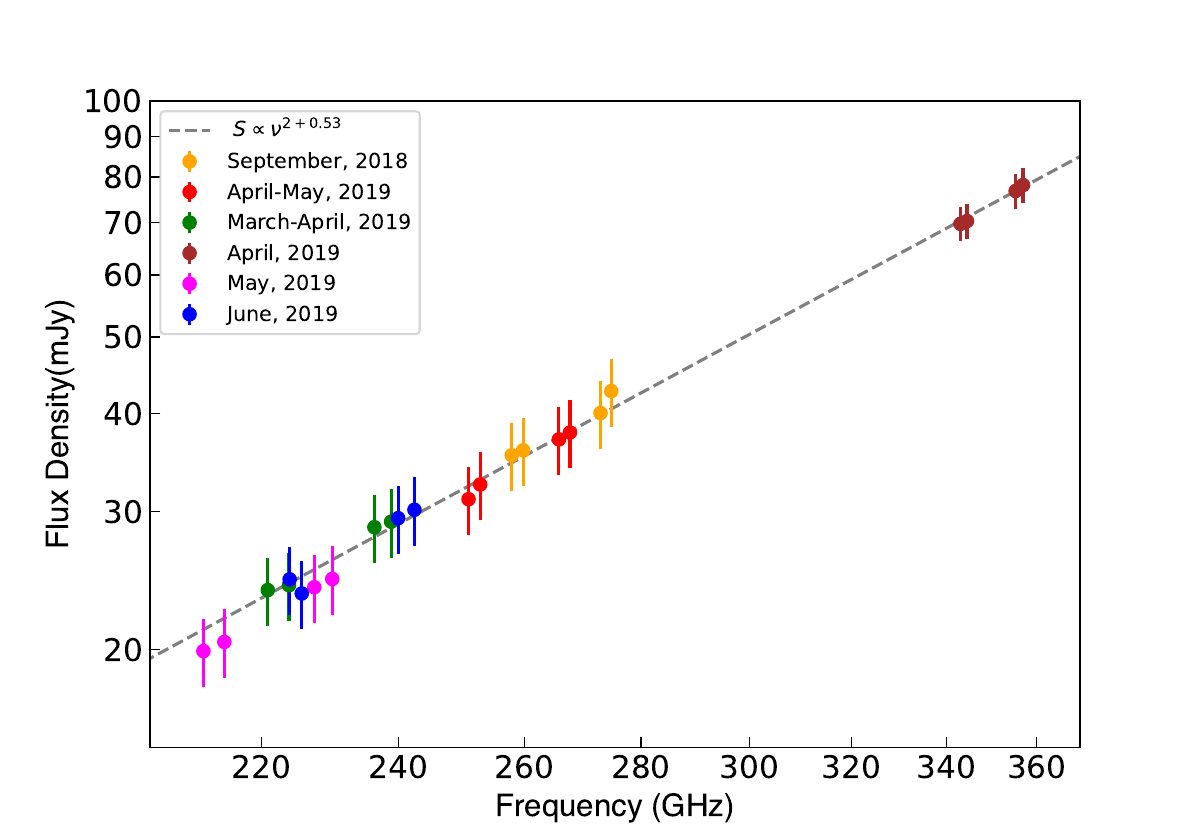}
			\caption{Spectral Energy Distribution of the ALMA continuum emission of Sakurai's object. The dashed line indicates a fit using a power law function, as described in the main text. The error bars indicate the nominal uncertainty in the absolute calibration of the ALMA observations of 5\%.}\label{fig:Fig2}
		\end{center} 
	\end{figure}
	
	In Fig.~\ref{fig:Fig1} we present an image of the millimetre (1.3~mm) continuum emission of Sakurai's object with an unprecedented angular-resolution of 20~milliarcsec (70~AU). The image reveals an emitting region consisting of a bright compact central component and faint extended structures elongated towards the northeast-southwest direction. The total flux at 233~GHz is 25~mJy and the peak of the emission is located at (J2000) R.A.=17$^{\rm h}$52$^{\rm m}$32$\rlap{.}^{\rm s}$6990 $\pm$ 0$\rlap{.}^{\rm s}$0002, Dec.=$-$17$^{\circ}$41$^{\prime}$7$\rlap{.}^{\prime\prime}$915 $\pm$ 0$\rlap{.}^{\prime\prime}$003. A Gaussian fit to the bright compact central component gives a deconvolved size of 44$\pm$3$\times$27$\pm$2~milliarcsec (154$\pm$10$\times$95$\pm$7 AU) with a P.A. of 90$^{\circ}$$\pm$6$^{\circ}$, shown as a grey ellipse in Fig~\ref{fig:Fig1}. The peak brightness temperature of this component is $T_\mathrm{B}$=217~K. The faint extended emission lies along P.A.$\approx$60$^{\circ}$ with a total length of approximately 230~milliarcsec (800~AU). Both the major axis of the bright compact central component and the axis of elongation of the faint extended structures are rotated compared to the bipolar outflow seen in near-infrared images, P.A.=21$^{\circ}$$\pm$5$^{\circ}$ \citep{Hinkle2020}, suggesting that the ALMA continuum is not tracing this bipolar outflow. In fact, the bright compact central component is oriented nearly perpendicularly to the bipolar outflow. Furthermore, its brightness temperature is in excellent agreement with the temperature ($T\approx200$~K) of the cold dust disc hinted at by mid-IR observations \citep{Chesneau2009,Hinkle2020,Evans2020}. This strongly suggests that the millimetre continuum emission of the bright compact central component is produced by such a cold dust disc\footnote{The structure referred to as a `disc' may also exhibit a torus-like morphology, but for consistency with the nomenclature in the literature, it will be referred to as a disc unless otherwise indicated.}, implying that our ALMA image is the first direct image of it.  
	
	Early analysis of IR observations indicated that the circumstellar dust of Sakurai's object primarily consisted of graphitic carbon \citep{Eyres1998,Tyne2002}. Subsequent studies using Spitzer Space Telescope and James Clerk Maxwell Telescope (JCMT) data suggested that the graphitic carbon fraction decreased, while the amorphous carbon fraction increased \citep{Evans2004,Evans2006}. \cite{Chesneau2009} noted the absence of spectral features in the MIDI IR spectrum, which they interpreted as indicative of the dominance of amorphous carbon grains in the dust composition. Furthermore, \cite{Evans2020} reported evidence for weak 6–7$\mu$m absorption, which they attribute to hydrogenated amorphous carbon formed in material ejected by Sakurai’s object at early stages after the born again event. 
	
	While all of this evidence appears to suggest that the dust is predominantly composed of amorphous carbon, there remains a need for additional constraints on the properties of the dust grains. Some of these constraints can be obtained from the analysis of data taken within the far-infrared to millimetre spectral range. In particular, at mm and sub-mm wavelengths, valuable insights into the characteristics of the emitting dust can be gained by assuming optically thin emission and fitting the observed flux densities to a power law of the form $S_{\nu} \propto \nu^{2+\beta}$, where $\beta$ represents the dust opacity index. For amorphous carbon, $\beta \approx 0.7-1$, whereas for graphitic carbon, $\beta \approx 2$ \cite[e.g.][]{Mennella1995,Mennella1998}. From our ALMA observations we obtained the continuum flux densities of Sakurai's object for the frequency range 212-357~GHz. The fluxes were measured by integrating a region containing emission exceeding 3 times the rms noise level of the image, and they are listed in Table \ref{Tab:2} and shown in Fig.~\ref{fig:Fig2}. A power law fit to the data gives an opacity index $\beta$=0.53$\pm$0.03. It is noteworthy that, up to this point, this value represents the most precisely determined dust opacity index for this source within the mm and sub-mm wavelengths range. This value is close to the one expected for the sample BE of amorphous carbon analogues studied by \cite{Mennella1998}, albeit being slightly smaller. The reason for this could be, as \cite{Evans2004} points out, that the emission is produced by large grains and/or or there has been an increase in the population of hotter grains (which have flatter emissivities). 
	
	Assuming isothermal dust with a temperature $T_{\rm dust}=217$~K (i.e. assuming that the peak continuum emission is optically thick) and that most of the dust emission is optically thin, the mass of the dust can be obtained using the following expression: 
	
	\begin{equation*}
		M_{\rm dust}=\frac{S_{\nu}\,D^{2}}{\kappa_{\nu}\,B_{\nu}(T_{\rm dust})}, 
	\end{equation*}
	
	where $S_{\nu}$ is the flux density of the continuum emission, $D$ is the distance to the source, $\kappa_{\nu}=\kappa_{0}(\nu/\nu_{0})^{\beta}$ is the dust absorption coefficient and $B_{\nu}(T)$ is the Planck function. Dust absorption coefficient of amorphous carbon in the literature range from $\kappa_{300\,{\rm GHz}} \sim 0.9 \, \text{cm}^{2}\,\text{g}^{-1}$ up to $\kappa_{300\,{\rm GHz}} \sim 90 \, \text{cm}^{2}\,\text{g}^{-1}$ \citep{Draine1984, Ossenkopf1994, Mennella1998, Suh2000}. Thus, using the dust opacity index obtained from our ALMA observations ($\beta = 0.53 \pm 0.03$, see Fig.~\ref{fig:Fig2}), the calculated mass of amorphous carbon dust in Sakurai's object is $6 \times 10^{-3}\,M_{\odot}$ and $6 \times 10^{-5}\,M_{\odot}$ for $\kappa_{300\,{\rm GHz}} \sim 0.9 \, \text{cm}^{2}\,\text{g}^{-1}$ and $\kappa_{300\,{\rm GHz}} \sim 90 \, \text{cm}^{2}\,\text{g}^{-1}$, respectively. Estimations of the dust mass based on mid-IR observations range from $1.8$ to $6 \times 10^{-5}\,M_{\odot}$ \citep{Chesneau2009,Hinkle2020,Evans2020}. Thus, in order to obtain a dust mass consistent with these previous estimations, it is necessary to consider amorphous carbon dust with a relatively large absorption coefficient in the sub-mm wavelength regime. 
	
	\begin{figure*}
		\begin{center} 
			\includegraphics[angle=0,width=\linewidth]{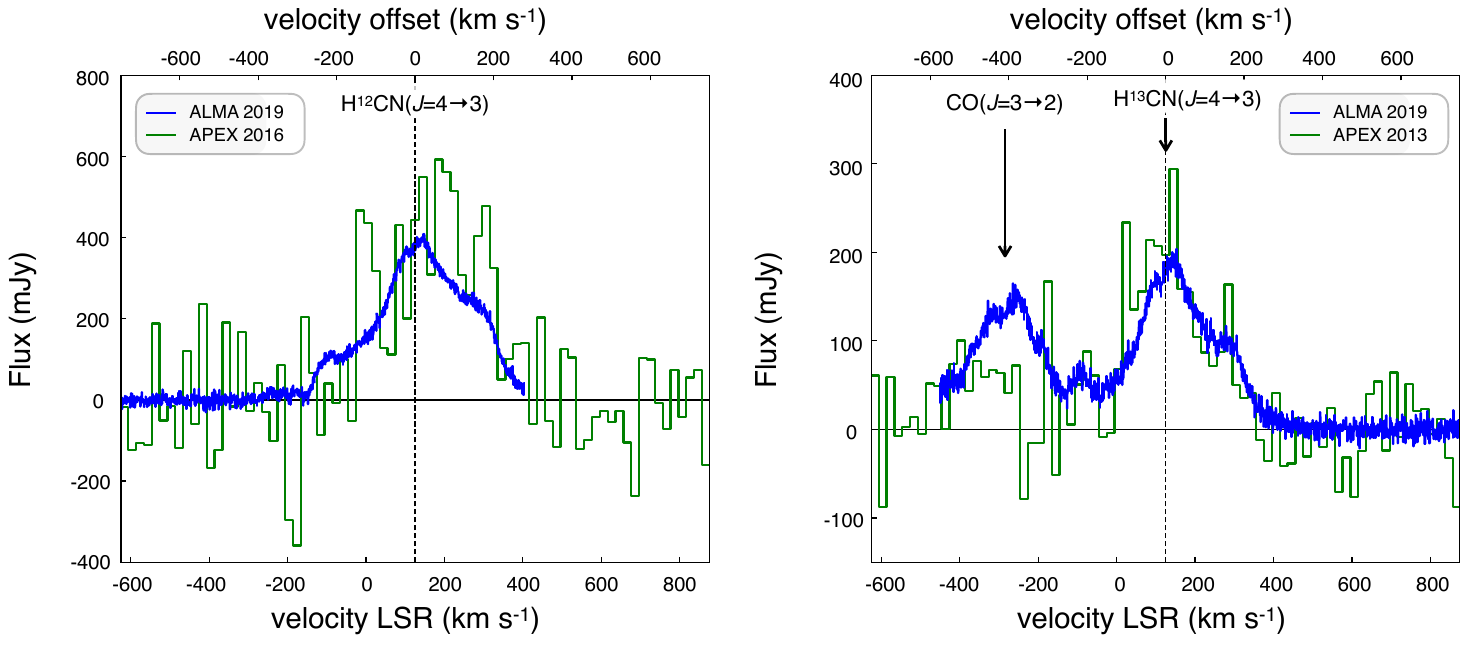}
			\caption{Sakurai's object H$^{12}$CN($J$=4$\rightarrow$3), H$^{13}$CN($J$=4$\rightarrow$3) and CO($J$=3$\rightarrow$2) line emission. The green line is the spectrum of the emission line detected previously with APEX. The blue line is the spectrum observed with ALMA. The spectral resolution of the APEX and ALMA observations is 20 km s$^{-1}$ and 0.8 km s$^{-1}$, respectively. The bottom axis indicates the local standard of rest (LSR) velocity and the top axis indicated the velocity offset from the systemic velocity assumed to be 125 km s$^{-1}$.}\label{fig:Fig3}
		\end{center} 
	\end{figure*}
	
	\begin{figure*}
		\begin{center} 
			\includegraphics[angle=0,width=\linewidth]{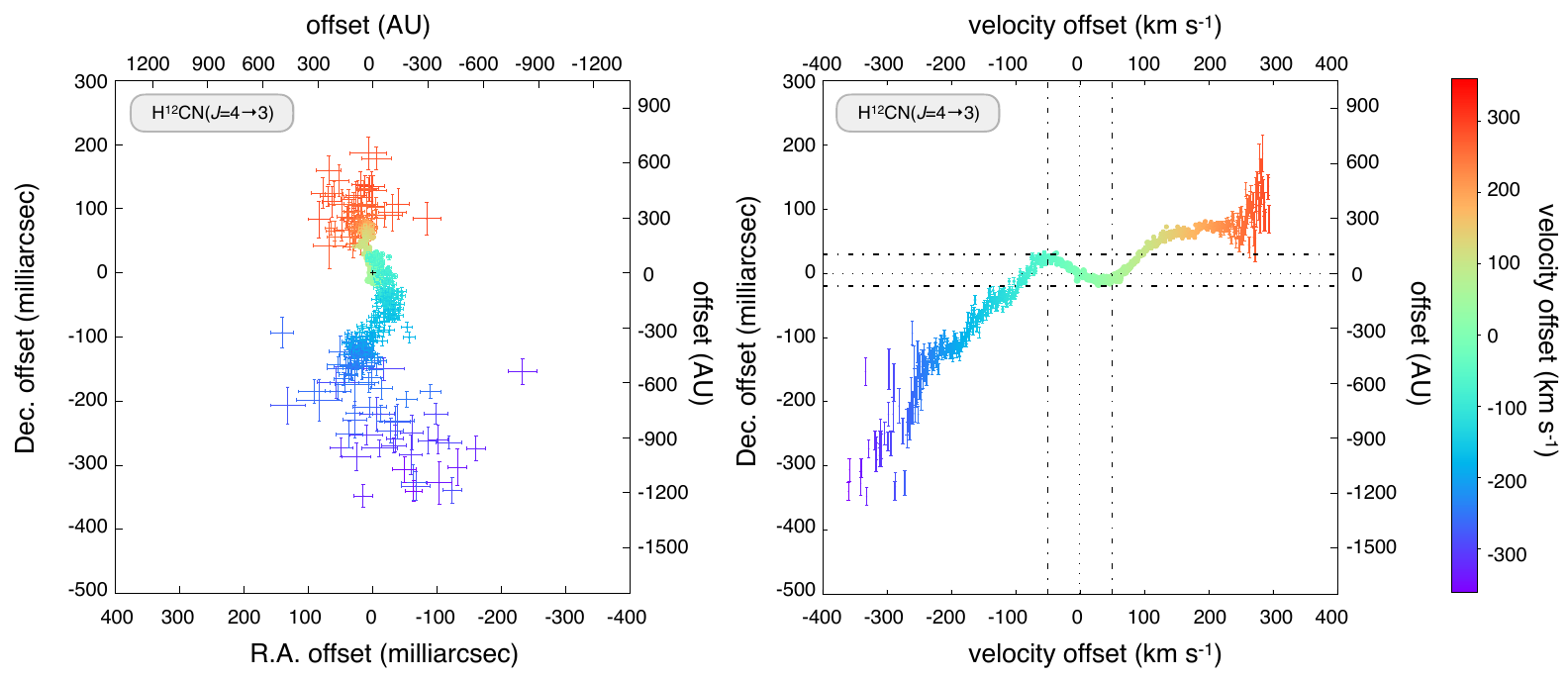}
			\caption{Spatial-kinematical distribution of the H$^{12}$CN($J$=4$\rightarrow$3) line emission in Sakurai's object. {\bf Left:} Spatial distribution of the centroid positions of the H$^{12}$CN($J$=4$\rightarrow$3) line emission at each individual velocity channel. {\bf Right:} Position-velocity diagram of the H$^{12}$CN($J$=4$\rightarrow$3) line emission centroids along the Declination axis. The vertical and horizontal dashed lines indicate the velocity offset and Declination offset, respectively, over which an inversion of the velocity gradient is observed. The error bars correspond to the uncertainties of the centroid positions. The axis indicating the linear scale assume a distance of 3.5 kpc to the source.}\label{fig:Fig4}
		\end{center} 
	\end{figure*}
	
	\begin{figure*}
		\begin{center} 
			\includegraphics[angle=0,width=\linewidth]{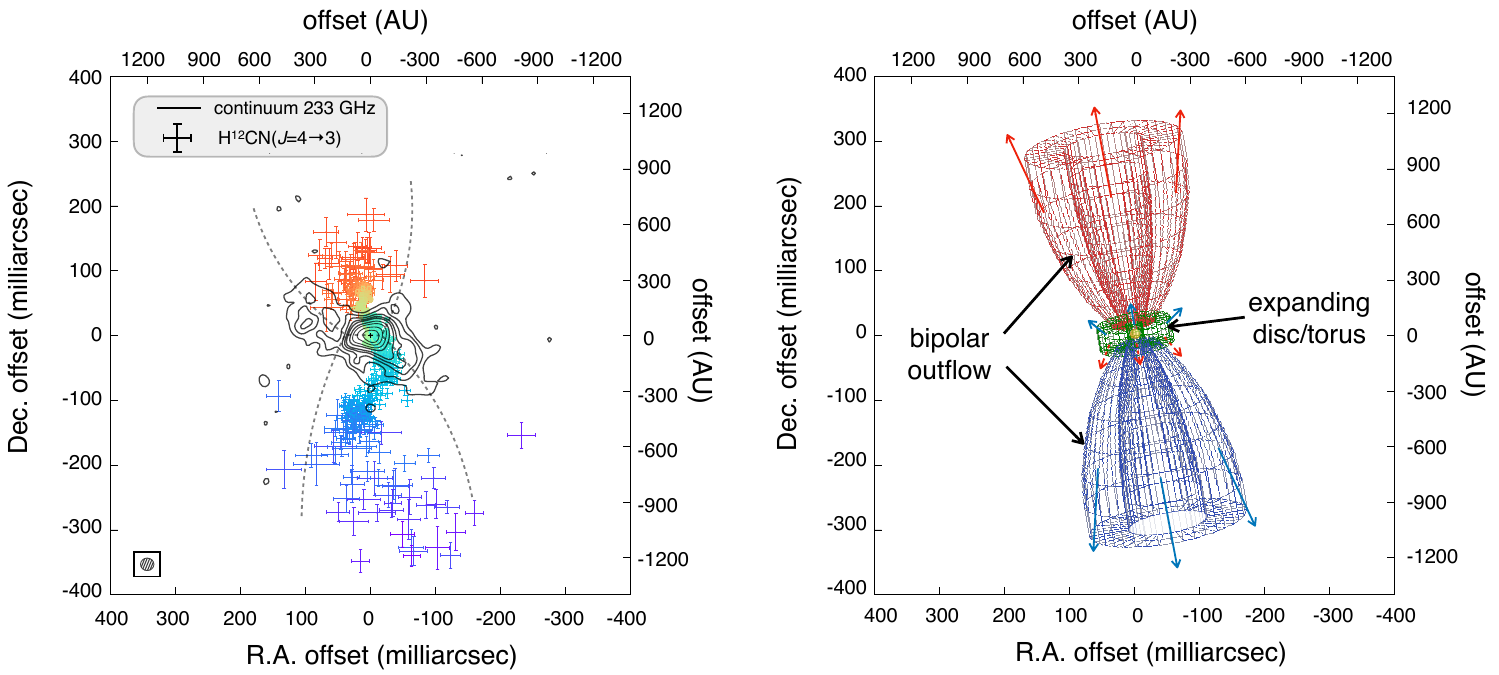}
			\caption{Bipolar outflow and expanding disc around Sakurai's object. {\bf Left:} H$^{12}$CN($J$=4$\rightarrow$3) line emission and 233~GHz continuum emission. The dashed lines delineate an hourglass morphology hinted at by the continuum emission. {\bf Right:} Hourglass model for the spatial-kinematical distribution of the bipolar outflow in Sakurai's object. The right and top axis indicating the linear scale assume a distance of 3.5~kpc to the source.}\label{fig:Fig5}
		\end{center} 
	\end{figure*}

	\subsection{Line emission: an expanding disc and a fast bipolar outflow in Sakurai's object}
	
	Carbon-rich molecules (CN, C$_{2}$, CO) were detected in the ejecta of Sakurai's object as early as one year after it was discovered to be undergoing the born-again event \citep{Eyres1998}. Subsequently, H-containing species (e.g. HCN and C$_{2}$H$_{2}$) were also found in the ejected material \citep{Evans2006}. All of these molecular species were detected against the dust continuum emission, probing primarily the molecular gas along the line-of-sight direction. More recently, \citet{Tafoya2017} detected for the first time emission of H$^{12}$CN($J$=4$\rightarrow$3), H$^{13}$CN($J$=4$\rightarrow$3), and H$^{12}$CN($J$=2$\rightarrow$1) lines, and \cite{vanHoof2018} presented semi-resolved ALMA images of the line of emission of CN, CO, and HC$_{3}$N. While these observations suggest that the CO molecules are situated in proximity to the dusty region \citep{Eyres_2004, Worters2009, vanHoof2018}, the HCN and CN are believed to reside in bipolar lobes \citep{Evans2006, vanHoof2018}. However, due to the limited angular resolution of previous observations, it has not been possible to discern the specific regions from which the molecular emission originates.
	
	To further investigate the molecular component of the ejecta around Sakurai's object, we conducted ALMA observations targeting the H$^{12}$CN($J$=4$\rightarrow$3), H$^{13}$CN($J$=4$\rightarrow$3), and CO($J$=3$\rightarrow$2) line emissions.
	These observations provided an angular resolution of approximately 300 milliarcsec ($\sim$1000 AU) and a spectral resolution of 0.8 km s$^{-1}$. Figure~\ref{fig:Fig3} shows the spectra obtained from the ALMA observations, alongside the spectra acquired using the Atacama Pathfinder Experiment (APEX) in 2013 and 2016 \citep{Tafoya2017}. Due to the lower S/N of the APEX spectra, accurately measuring any changes in the brightness of the H$^{12}$CN($J$=4$\rightarrow$3) and H$^{13}$CN($J$=4$\rightarrow$3) lines between the APEX and ALMA observations is challenging. However, the spectrum in the right panel of Fig.~\ref{fig:Fig3} clearly shows that the CO($J$=3$\rightarrow$2) line, observed in the ALMA data, was either not detected or was significantly weaker in the APEX observations. This could be the result of either the recent formation of significant amounts of CO molecules or changes in the physical conditions of the emitting gas. We note that although CO molecules were found in the newly formed ejecta of Sakurai's object \citep{Eyres1998, Pavlenko2004}, they were detected through absorption features in the near-infrared spectra, which is sensitive to even small amounts of CO. Consequently, the cause of the observed enhancement in the CO($J$=3$\rightarrow$2) line intensity remains uncertain.
	
	The widths of the molecular lines are notably broad. The velocity offset range covered by the H$^{12}$CN($J$=4$\rightarrow$3) line extends from $-$350~km~s$^{-1}$ to 300~km~s$^{-1}$, with the red-shifted portion even exceeding the bandwidth of our spectral window. In contrast, the H$^{13}$CN($J$=4$\rightarrow$3) and CO($J$=3$\rightarrow$2) lines are blended, preventing us from spatially separating their emissions. Furthermore, the blue-shifted part of the CO($J$=3$\rightarrow$2) line also extends beyond the limits of our spectral window. Despite these limitations, the significantly higher spectral resolution of the ALMA observations enables the identification of distinct velocity components for each molecule. Additionally, the achieved high S/N allows for precise determination of the centroid of the molecular emission in each velocity channel, surpassing the size of the synthesised beam \citep{Condon1997}. Specifically, the uncertainties of the centroid positions are obtained as $\sigma_{\rm centroid}$=$({\sigma_{\theta}^{2}+\sigma_{\rm BP}^{2}})^{1/2}$, where $\sigma_{\theta}$$\approx$$\theta_{\rm beam}$/(2 SNR) and $\sigma_{\rm BP}$=$\theta_{\rm beam}$($\sigma_{\phi}$/360$^{\circ}$). In these expressions $\theta_{\rm beam}$ is the beam size and $\sigma_{\phi}$ is phase noise of the bandpass calibrator \citep[e.g.][]{Zhang2017}. For our observations, $\theta_{\rm beam}$=300 milliarcsecs and $\sigma_{\phi}$=0.08$^{\circ}$.
	
	In Fig.~\ref{fig:Fig4} we present the spatial distribution and a position-velocity (PV) diagram of the H$^{12}$CN($J$=4$\rightarrow$3) line emission. The emission exhibits a clear north-south velocity gradient, with the blue-shifted and red-shifted components positioned to the south and north, respectively, of the peak of the continuum emission. The spatial distributions and PV diagrams of the H$^{13}$CN($J$=4$\rightarrow$3) and CO($J$=3$\rightarrow$2) lines show similar characteristics to those of the H$^{12}$CN($J$=4$\rightarrow$3) line, but they emitting region seems to be more compact. This aligns perfectly with the direction and velocity gradient of the near-infrared outflow \citep{Hinkle2020}, confirming that the H$^{12}$CN($J$=4$\rightarrow$3) line emission indeed traces the bipolar outflow of Sakurai's object.
	
	Upon closer examination of the PV diagram, it is evident that there is a distinctive change in the velocity gradient. In the central region, where velocity offsets range from $-$50 to 50 km~s$^{-1}$ (see right panel of Fig.~\ref{fig:Fig4}), the slope is inverted. The horizontal dashed lines in the PV diagram indicate that this emission extends over approximately 50~milliarcsec ($\sim$175 AU), which is comparable to the size of the dust disc observed in the ALMA continuum image. This particular pattern of an inverted velocity gradient is commonly observed in systems that exhibit both a bipolar outflow and an expanding equatorial disc \citep[e.g.][]{Alcolea2007}. Thus, it is likely that the inverted velocity gradient is produced by the expansion of the equatorial disc seen in the millimetre continuum image (see Fig.\ref{fig:Fig1}).
	
	For an expanding circular thin disc of radius $R$, whose polar axis has an inclination $i$ with respect to the line of sight, that is $i$=0$^{\circ}$ and $i$=90$^{\circ}$ correspond to the disc seen face-on and edge-on, respectively, the projected minor axis in angular units, $\theta_{\rm disc}$, and its line-of-sight velocity, $v_{\rm los}$, are given by: $\theta_{\rm disc}=2R{\rm cos}\,i/D$ and $v_{\rm los}=v_{\rm exp}{\rm sin}\,i$, respectively, where $D$ is the distance to the source and $v_{\rm exp}$ is the expansion velocity of the disc. Assuming a constant expansion velocity, the radius of the disc can be calculated as $R=v_{\rm exp}\,\tau$, where $\tau$ is the time since the ejection of the molecular gas. Thus, the inclination of the disc can be obtained from the following expression: 
	
	\begin{equation*}\label{Eq:1}
		{\rm tan}\,i= 3.0 \left[\frac{v_{\rm los}}{\rm 50\,km\,s^{-1}}\right]\left[\frac{\tau}{\rm 25\,years}\right]\left[\frac{\theta_{\rm disc}}{\rm 50\,milliarcsec}\right]^{-1}\left[\frac{D}{\rm 3.5\,kpc}\right]^{-1}. 
	\end{equation*}
	
	From our observations we find that $v_{\rm los}$=50~km~s$^{-1}$ and $\theta_{d}$=50~milliarcsec, and considering that the ejection of the material occurred around 25~years ago, the resulting inclination of the disc is $i$=72$^{\circ}$. This result confirms that the equatorial disc is seen almost edge-on, completely obscuring the central star \citep{Chesneau2009,Hinkle2020}. The de-projected expansion velocity and the radius of the disc are $v_{\rm exp}$=53~km~s$^{-1}$ and $R$=277~AU, respectively. 
	
	We note that the derived inclination of the disc from our ALMA observations is opposite to the one deduced from the VLTI observations \citep{Chesneau2009}. While the VLTI observations suggest that the northern side of the disc is
	tipped toward us, which would mean that the northern edge of the expanding disc is red-shifted, the ALMA observations yield the opposite for the inner 50 milliarcsec: blue-shifted emission toward the north and red-shifted emission toward the south (cf. right panel of Fig.~\ref{fig:Fig4}). Furthermore, the inclination obtained from the VLTI observations implies that the northern part of the outflow should be red-shifted, which is opposite to the velocity gradient seen from the near-infrared and ALMA observations. Thus, we conclude that the southern side of the disc is the one tipped toward us.

	The derived inclination provides an estimate of the de-projected expansion velocity of the bipolar outflow detected through the H$^{12}$CN($J$=4$\rightarrow$3) emission, which is approximately 1000~km~s$^{-1}$, assuming a fairly collimated structure. Such high expansion velocities for the bipolar outflow of Sakurai's object are not unexpected. For instance, when correcting for the inclination, the expansion velocity of the He I 10830 $\AA$ emission line observed with the Gemini NIFS shows a similar value. Also, \cite{Kerber2002} concluded that the bipolar outflow, traced by the [N II] $\lambda$6583 line emission, exhibited an expansion velocity of about 800~km~s$^{-1}$. It is worth noting that such a high expansion velocity implies that the observed lobes should be more than four times larger than observed, assuming that the molecular bipolar outflow formed at the same time as the rest of the material was ejected. This could indicate that the H$^{12}$CN($J$=4$\rightarrow$3) emission does not fully trace the entirety of the molecular outflow but only the regions where the excitation conditions required to observe HCN molecules are optimal. On the other hand, we cannot dismiss the possibility that the outflow traced by the H$^{12}$CN($J$=4$\rightarrow$3) emission has a relatively wide opening angle. In such case, the expansion velocity of the molecular material would be lower, which could explain the smaller observed size of the molecular lobes.

	The left panel of Fig.~\ref{fig:Fig5} shows the continuum emission superimposed on the spatial distribution of the H$^{12}$CN ($J$=4$\rightarrow$3) line emission. Towards the north, the continuum emission shows two protrusions that appear to be tracing the base of an hourglass-shaped outflow (indicated by the dashed lines). On the southern side, the continuum emission exhibits another such protrusion. A sketch of the inferred geometry and orientation of the material around Sakurai's object is shown in the right panel of Fig.~\ref{fig:Fig5}. As suggested by the protrusions in the continuum emission, the bipolar outflow lobes have an opening angle of around 60$^{\circ}$ and their walls seem to be denser than the regions close to the polar axis. In addition, as mentioned above, the spatio-kinematical configuration of the molecular emission is the same as the one derived from observations of the He I 10830\AA\, and the [C I] 9850\AA\, lines \citep{Hinkle2020}. Thus, it is possible that the expanding atomic gas forms a more collimated outflow that entrains the surrounding molecular material around it, forging the molecular outflow.

	\section{Concluding remarks}
	\label{sec:discussion}
	
	It is remarkable that most known objects that are thought to have experienced a born-again event (A30, A78, V605 Aql, Sakurai's object), have a strikingly similar morphology, namely an expanding equatorial disc and a bipolar outflow \citep{Toala2015,Hinkle2020,Tafoya2022,RodriguezGonzalez2022}. The formation of this particular morphology has most often been attributed to a stellar/sub-stellar binary companion. For example A30, a common-envelope phase following the born-again event has been proposed to explain the observed physical structure and abundances of the ejecta around the central star \citep{RodriguezGonzalez2022}. \citet{Tafoya2022} proposed a similar common-envelope scenario to account for the expanding equatorial structure and bipolar outflow seen in their ALMA observations of V605 Aql. In contrast, standard models of the born-again event are based on single-star evolution that would result in a simple, spherical morphology of the ejecta. This has so far been observed in only one case, HuBi 1 \citep{Toala2021b,Rechy-Garcia2020}. It is difficult to explain the more complex morphology observed for the other observed objects without invoking common envelope evolution, which would necessitate a paradigm shift since close-binary interactions would certainly influence the physics of the born-again process.

	Our ALMA observations have revealed, in unprecedented detail, the morpho-kinematical structure of the material surrounding the youngest born-again star, allowing us to study the initial stages of the ejection process. These results set stringent observational constraints for future models that will need to be consistent not only with the observed chemistry, but also the physical structure and intriguing morphology observed in the different stages of the born-again process.

	\begin{acknowledgements}
		This paper makes use of the following ALMA data: ADS / JAO.ALMA \# 2017.1. 00017.S, 2018.1.00088.S and 2018.1.00341.S. ALMA is a partnership of ESO (representing its member states), NSF (USA) and NINS (Japan), together with NRC (Canada), MOST and ASIAA (Taiwan), and KASI (Republic of Korea), in cooperation with the Republic of Chile. The Joint ALMA Observatory is operated by ESO, AUI/NRAO and NAOJ. DT acknowledges support from Onsala Space Observatory for the provisioning of its facilities support. The Onsala Space Observatory national research infrastructure is funded through Swedish Research  Council grant No 2017-00648.
	\end{acknowledgements}
	
	%
	%
	
	\bibliographystyle{aa} 

	\begin{appendix} 
		
			\onecolumn
			\begin{landscape}
				\section{ALMA Observations}
				\begin{longtable}{r r r r c c r}
					\caption{Parameters of the ALMA observations}\label{Tab:1}\\    
					\hline
					\multicolumn{1}{c}{Observation date}& \multicolumn{1}{c}{Frequency$^{\rm a}$} & \multicolumn{1}{c}{Flux \& Bandpass calibrator} & \multicolumn{1}{c}{Phase calibrator}& \multicolumn{1}{c}{Bandwidth$^{\rm b}$} & \multicolumn{1}{c}{rms$^{\rm c}$} & \multicolumn{1}{c}{Beam} \\ 
					\multicolumn{1}{c}{(YYYY-MM-DD)} & \multicolumn{1}{c}{(GHz)} & \multicolumn{1}{c}{(name, Jy)} & \multicolumn{1}{c}{(name, mJy)} & \multicolumn{1}{c}{(GHz)} & \multicolumn{1}{c}{($\mu$Jy beam$^{-1}$)} & \multicolumn{1}{c}{(mas$\times$mas,degree)$^{\rm c}$} \\ \hline
					\endfirsthead
					\multicolumn{7}{c} {\tablename\ \thetable\ -- \textit{Continued from previous page}} \\
					\hline
					\multicolumn{1}{c}{Observation date}& \multicolumn{1}{c}{Frequency$^{\rm a}$} & \multicolumn{1}{c}{Flux \& Bandpass calibrator} & \multicolumn{1}{c}{Phase calibrator}& \multicolumn{1}{c}{Bandwidth$^{\rm b}$} & \multicolumn{1}{c}{rms$^{\rm c}$} & \multicolumn{1}{c}{Beam} \\ 
					\multicolumn{1}{c}{(YYYY-MM-DD)} & \multicolumn{1}{c}{(GHz)} & \multicolumn{1}{c}{(name, Jy)} & \multicolumn{1}{c}{(name, mJy)} & \multicolumn{1}{c}{(GHz)} & \multicolumn{1}{c}{($\mu$Jy beam$^{-1}$)} & \multicolumn{1}{c}{(mas$\times$mas,degree)$^{\rm c}$} \\ \hline
					\endhead
					\hline \multicolumn{7}{c}{\textit{Continued on next page}} \\
					\endfoot
					\multicolumn{7}{l}{$^{\rm a}$Central frequency of the spectral window.}\\
					\multicolumn{7}{l}{$^{\rm b}$Effective bandwidth of the spectral window used to produce the continuum images.}\\
					\multicolumn{7}{l}{$^{\rm c}$rms noise over the effective bandwidth used to produce the continuum images.}\\
					\multicolumn{7}{l}{$^{\rm d}$mas=milliarcsec.}\\
					\multicolumn{7}{l}{$^{\rm e}$Data taken at two different dates were used to produce the continuum images.}
					\endlastfoot
					\multicolumn{7}{c}{Project code: 2017.1.00017.S}\\\hline
					2018-09-28& 257.049& J1924-2914, 2.977 & J1743-1658, 142.5$\pm$0.8&1.86 & 130& 268$\times$218, 69.7 \\ 
					& 258.911& 2.963 & 142.2$\pm$0.8& 0.91& 130& 264$\times$214, 69.6\\ 
					& 272.041& 2.868 & 138.6$\pm$0.8& 0.82& 160& 252$\times$204, 68.8\\ 
					& 273.904& 2.855 & 139.1$\pm$0.7& 1.64& 160& 251$\times$204, 68.6\\ \hline
					\multicolumn{7}{c}{Project code: 2018.1.00088.S}\\\hline
					2019-05-01& 212.043& J1924-2914, 2.746 & J1743-1658, 143.4$\pm$0.9& 1.52& 31& 515$\times$454, $-$79.9 \\ 
					& 214.933& 2.723 & 142.8$\pm$1.0& 1.86& 30& 514$\times$452, $-$81.1\\ 
					& 227.357& 2.634 & 139.6$\pm$1.0& 0.89& 37& 487$\times$431, $-$81.4\\ 
					& 230.038& 2.616 & 138.6$\pm$1.0& 0.86& 40& 482$\times$425, $-$81.4\\ 
					2019-03-31[2019-04-11]$^{\rm d}$& 220.945& J1924-2914[J1517-2422], 2.900[2.453]& J1743-1658, 157.0$\pm$0.8[145.0$\pm$0.5]& 1.16& 48& 792$\times$571, $-$88.1\\ 
					& 223.944& J1924-2914[J1517-2422], 2.875[2.443]& 155.9$\pm$0.8[144.3$\pm$0.5]& 1.85& 41& 780$\times$560, $-$88.5\\ 
					& 236.454& J1924-2914[J1517-2422], 2.776[2.402]& 151.9$\pm$0.9[140.0$\pm$0.6]& 1.21& 54& 745$\times$533, $-$88.9\\ 
					& 238.941& J1924-2914[J1517-2422], 2.757[2.394]& 151.2$\pm$0.7[139.5$\pm$0.6]& 1.85& 52& 735$\times$531, $-$88.9\\ 
					2019-04-30[2019-05-01]$^{\rm d}$& 250.982& J1924-2914, 2.483& J1743-1658, 130.8$\pm$0.9[126.4$\pm$0.9]& 1.25& 57& 492$\times$403, 89.2\\ 
					& 252.850& 2.472& 130.2$\pm$0.9[126.5$\pm$1.0]& 1.41& 57& 486$\times$400, 88.9\\ 
					& 265.808& 2.400& 126.9$\pm$0.7[123.7$\pm$1.0]& 1.08& 70& 472$\times$412, $-$66.9\\ 
					& 267.677& 2.390& 126.4$\pm$0.7[122.9$\pm$0.7]& 1.19& 71& 458$\times$381, 89.8\\ 
					2019-06-22& 224.000& J1924-2914, 2.913 & J1743-1658, 157.0$\pm$1.0& 1.75& 73& 25$\times$24, 48.2 \\ 
					& 226.000& 2.898 & 156.1$\pm$0.6& 1.48& 73& 24$\times$23, $-$82.8\\ 
					& 240.000& 2.795 & 152.6$\pm$0.8& 1.75& 77& 23$\times$22, $-$89.4\\ 
					& 242.000& 2.782 & 151.8$\pm$0.7& 1.75& 84& 23$\times$21, $-$87.2\\ \hline
					\multicolumn{7}{c}{Project code: 2018.1.00341.S}\\\hline
					2019-04-21& 343.009& J1924-2914, 2.051 & J1742-1517, 149.2$\pm$0.4& 1.37& 240& 335$\times$303, 77.3 \\ 
					& 344.967& 2.044 & 148.7$\pm$0.4& 0.79& 260& 336$\times$309, 72.0\\ 
					& 355.009& 2.008 & 147.8$\pm$0.5& 0.76& 290& 326$\times$298, 73.8\\ 
					& 356.897& 2.002 & 147.1$\pm$0.5& 1.36& 280& 323$\times$293, 76.8\\ \hline
				\end{longtable}
			\end{landscape}
			
			\begin{table}[h]
				\centering
				\caption{Frequency and Flux Density Measurements}\label{Tab:2}
				\begin{tabular}{cc}
					\hline
					Frequency (GHz)\tablefootmark{a} & Flux Density (mJy)\tablefootmark{b} \\
					\hline
					212.071 & 19.91 \\
					214.908 & 20.45 \\
					220.918 & 23.81 \\
					223.913 & 24.15 \\
					224.006 & 24.57 \\
					225.733 & 23.57 \\
					227.537 & 24.01 \\
					230.150 & 24.60 \\
					236.392 & 28.63 \\
					238.911 & 29.10 \\
					239.974 & 29.40 \\
					242.482 & 30.13 \\
					250.953 & 31.08 \\
					252.820 & 32.45 \\
					257.927 & 35.36 \\
					259.805 & 35.88 \\
					265.775 & 37.04 \\
					267.645 & 37.80 \\
					272.886 & 40.02 \\
					274.756 & 42.68 \\
					342.960 & 69.70 \\
					344.387 & 70.27 \\
					355.186 & 76.75 \\
					356.856 & 78.11 \\
					\hline
				\end{tabular}
				\tablefoot{
					\tablefoottext{a}{Central frequency of the bandwidth used to create the continuum images.}\\
					\tablefoottext{b}{Flux density obtained by integrating a region containing emission exceeding 3 times the rms noise level of the image.The nominal uncertainty in the absolute calibration of the ALMA observations of 10\% is adopted.}
				}
			\end{table}
			
		\end{appendix}
		
	\end{document}